\documentclass[%
	manyauthors,
  nocleardouble,
  COMPASS,
  american]{cernphprep}

\usepackage[loadSvnInfoPackage=true]{packages}

\usepackage{macros}

\graphicspath{%
  {./plots/},%
  {./}%
}

\makeatletter
\g@addto@macro\bfseries{\boldmath}
\makeatother

\begin{document}

\begin{titlepage}
  \PHnumber{2015--015}
  \PHdate{17 July 2015}

  %
  %
  %
%
%

\title{Observation of a new narrow axial-vector meson $\PaOne[1420]$}
   \ShortTitle{Observation of a new narrow axial-vector meson $\PaOne[1420]$}
  \Collaboration{The COMPASS Collaboration}
  \ShortAuthor{The COMPASS Collaboration}

  %
  %
  %
%
%

\begin{abstract}
  The COMPASS collaboration at CERN has measured diffractive
  dissociation of \SI{190}{\GeVc} pions into the \threePi final state
  using a stationary hydrogen target.  A partial-wave analysis (PWA)
  was performed in bins of $3\pi$ mass and four-momentum transfer
  using the isobar model and the so far largest PWA model consisting
  of 88~waves.  A narrow $\JPC = 1^{++}$ signal is observed in the
  $\PfZero\,\pi$ channel.  We present a resonance-model study of a
  subset of the spin-density matrix selecting $3\pi$ states with
  $\JPC = \twoPP$ and \fourPP decaying into $\Prho\,\pi$ and with
  $\JPC = \onePP$ decaying into $\PfZero\,\pi$.  We identify a new
  \PaOne* meson with mass \SIaerr{1414}{15}{13}{\MeVcc} and width
  \SIaerr{153}{8}{23}{\MeVcc}.  Within the final states investigated
  in our analysis, we observe the new $\PaOne[1420]$ decaying only
  into $\PfZero\,\pi$, suggesting its exotic nature.
\end{abstract}
   \vspace*{60pt}
  \begin{flushleft}
  \end{flushleft}
  \vfill
  \Submitted{(to be submitted to Physical Review Letters)}
\end{titlepage}

{\pagestyle{empty} %
%
%
\section*{The COMPASS Collaboration}
\label{app:collab}
\renewcommand\labelenumi{\textsuperscript{\theenumi}~}
\renewcommand\theenumi{\arabic{enumi}}
\begin{flushleft}
C.~Adolph\Irefn{erlangen},
R.~Akhunzyanov\Irefn{dubna}, 
M.G.~Alexeev\Irefn{turin_u},
G.D.~Alexeev\Irefn{dubna}, 
A.~Amoroso\Irefnn{turin_u}{turin_i},
V.~Andrieux\Irefn{saclay},
V.~Anosov\Irefn{dubna}, 
A.~Austregesilo\Irefnn{cern}{munichtu},
C.~Azevedo\Irefn{aveiro},           
B.~Bade{\l}ek\Irefn{warsawu}
% ...
\end{flushleft}

%
%
\begin{Authlist}
\item \Idef{turin_p}{University of Eastern Piedmont, 15100 Alessandria, Italy}
\item \Idef{aveiro}{University of Aveiro, Department of Physics, 3810-193 Aveiro, Portugal}
\item \Idef{bochum}{Universit\"at Bochum, Institut f\"ur Experimentalphysik, 44780 Bochum, Germany\Arefs{l}\Arefs{s}}
\item \Idef{bonniskp}{Universit\"at Bonn, Helmholtz-Institut f\"ur  Strahlen- und Kernphysik, 53115 Bonn, Germany\Arefs{l}}
\item \Idef{bonnpi}{Universit\"at Bonn, Physikalisches Institut, 53115 Bonn, Germany\Arefs{l}}
\item \Idef{brno}{Institute of Scientific Instruments, AS CR, 61264 Brno, Czech Republic\Arefs{m}}
\item \Idef{calcutta}{Matrivani Institute of Experimental Research \& Education, Calcutta-700 030, India\Arefs{n}}
\item \Idef{dubna}{Joint Institute for Nuclear Research, 141980 Dubna, Moscow region, Russia\Arefs{o}}
\item \Idef{erlangen}{Universit\"at Erlangen--N\"urnberg, Physikalisches Institut, 91054 Erlangen, Germany\Arefs{l}}
\item \Idef{freiburg}{Universit\"at Freiburg, Physikalisches Institut, 79104 Freiburg, Germany\Arefs{l}\Arefs{s}}
\item \Idef{cern}{CERN, 1211 Geneva 23, Switzerland}
\item \Idef{liberec}{Technical University in Liberec, 46117 Liberec, Czech Republic\Arefs{m}}
\item \Idef{lisbon}{LIP, 1000-149 Lisbon, Portugal\Arefs{p}}
\item \Idef{mainz}{Universit\"at Mainz, Institut f\"ur Kernphysik, 55099 Mainz, Germany\Arefs{l}}
\item \Idef{miyazaki}{University of Miyazaki, Miyazaki 889-2192, Japan\Arefs{q}}
\item \Idef{moscowlpi}{Lebedev Physical Institute, 119991 Moscow, Russia}
\item \Idef{munichtu}{Technische Universit\"at M\"unchen, Physik Department, 85748 Garching, Germany\Arefs{l}\Arefs{r}}
\item \Idef{nagoya}{Nagoya University, 464 Nagoya, Japan\Arefs{q}}
\item \Idef{praguecu}{Charles University in Prague, Faculty of Mathematics and Physics, 18000 Prague, Czech Republic\Arefs{m}}
\item \Idef{praguectu}{Czech Technical University in Prague, 16636 Prague, Czech Republic\Arefs{m}}
\item \Idef{protvino}{State Scientific Center Institute for High Energy Physics of National Research Center `Kurchatov Institute', 142281 Protvino, Russia}
\item \Idef{saclay}{CEA IRFU/SPhN Saclay, 91191 Gif-sur-Yvette, France\Arefs{s}}
\item \Idef{taipei}{Academia Sinica, Institute of Physics, Taipei, 11529 Taiwan}
\item \Idef{telaviv}{Tel Aviv University, School of Physics and Astronomy, 69978 Tel Aviv, Israel\Arefs{t}}
\item \Idef{triest_u}{University of Trieste, Department of Physics, 34127 Trieste, Italy}
\item \Idef{triest_i}{Trieste Section of INFN, 34127 Trieste, Italy}
\item \Idef{triest_ictp}{Abdus Salam ICTP, 34151 Trieste, Italy}
\item \Idef{turin_u}{University of Turin, Department of Physics, 10125 Turin, Italy}
\item \Idef{turin_i}{Torino Section of INFN, 10125 Turin, Italy}
\item \Idef{illinois}{University of Illinois at Urbana-Champaign, Department of Physics, Urbana, IL 61801-3080, U.S.A.}
\item \Idef{warsaw}{National Centre for Nuclear Research, 00-681 Warsaw, Poland\Arefs{u} }
\item \Idef{warsawu}{University of Warsaw, Faculty of Physics, 02-093 Warsaw, Poland\Arefs{u} }
\item \Idef{warsawtu}{Warsaw University of Technology, Institute of Radioelectronics, 00-665 Warsaw, Poland\Arefs{u} }
\item \Idef{yamagata}{Yamagata University, Yamagata, 992-8510 Japan\Arefs{q} }
\end{Authlist}

%
%
\vspace*{\baselineskip}
\renewcommand\theenumi{\alph{enumi}}
\begin{Authlist}
\item \Adef{a}{Also at Instituto Superior T\'ecnico, Universidade de Lisboa, Lisbon, Portugal}
\item \Adef{b}{Also at Department of Physics, Pusan National University, Busan 609-735, Republic of Korea and at Physics Department, Brookhaven National Laboratory, Upton, NY 11973, U.S.A. }
\item \Adef{c}{Supported by the DFG Research Training Group Programme 1102  ``Physics at Hadron Accelerators''}
\item \Adef{d}{Also at Chubu University, Kasugai, Aichi, 487-8501 Japan\Arefs{q}}
\item \Adef{e}{Also at KEK, 1-1 Oho, Tsukuba, Ibaraki, 305-0801 Japan}
\item \Adef{f}{Present address: Universit\"at Bonn, Helmholtz-Institut f\"ur Strahlen- und Kernphysik, 53115 Bonn, Germany}
\item \Adef{g}{Also at Moscow Institute of Physics and Technology, Moscow Region, 141700, Russia}
\item \Adef{h}{present address: RWTH Aachen University, III. Physikalisches Institut, 52056 Aachen, Germany}
\item \Adef{i}{present address: Uppsala University, Box 516, SE-75120 Uppsala, Sweden}
\item \Adef{l}{Supported by the German Bundesministerium f\"ur Bildung und Forschung}
\item \Adef{m}{Supported by Czech Republic MEYS Grant LG13031}
\item \Adef{n}{Supported by SAIL (CSR), Govt.\ of India}
\item \Adef{o}{Supported by CERN-RFBR Grant 12-02-91500}
\item \Adef{p}{\raggedright Supported by the Portuguese FCT - Funda\c{c}\~{a}o para a Ci\^{e}ncia e Tecnologia, COMPETE and QREN, Grants CERN/FP/109323/2009, CERN/FP/116376/2010 and CERN/FP/123600/2011}
\item \Adef{q}{Supported by the MEXT and the JSPS under the Grants No.18002006, No.20540299 and No.18540281; Daiko Foundation and Yamada Foundation}
\item \Adef{r}{Supported by the DFG cluster of excellence `Origin and Structure of the Universe' (www.universe-cluster.de)}
\item \Adef{s}{Supported by EU FP7 (HadronPhysics3, Grant Agreement number 283286)}
\item \Adef{t}{Supported by the Israel Science Foundation, founded by the Israel Academy of Sciences and Humanities}
\item \Adef{u}{Supported by the Polish NCN Grant DEC-2011/01/M/ST2/02350}
\item [{\makebox[2mm][l]{\textsuperscript{*}}}] Deceased
\end{Authlist}

 %
}
\newpage
\setcounter{page}{1}

\bibliographystyle{utphys_bgrube}
%
%
%

\setcounter{page}{1}

%
%
%

One of the great challenges in particle physics is the understanding
of how hadronic matter is constructed from its basic building blocks,
quarks and gluons. Although many possibilities are allowed, which all
follow the principle of confinement, almost all hadrons observed can
be described by the constituent quark model.  The known light-meson
spectrum is presently interpreted in terms of \qqbar quark-model
states that are associated with flavor SU(3) multiplets according to
their mass and \JPC quantum numbers.  For some \JPC combinations, more
states were reported than can be accommodated by SU(3) symmetry.
Depending on their coupling to specific production mechanisms and
their decay pattern, these states are interpreted as either carrying a
strong glueball component, \eg~\PfZero[1500], as molecular-type
excitations, \eg~\PfOne[1420], or as tetra-quark states.  For a
detailed review see \refCite{Klempt:2007cp}.  However, their exotic
structure is under debate, unlike for states that carry spin-exotic
quantum numbers, \eg $\JPC = \oneMP$, and hence cannot be \qqbar
states.  This is in contrast to the sector of heavy mesons containing
$c$~or $b$~quarks, where exotic mesons have clearly been identified,
\eg $X$, $Y$, and $Z$-states.  In particular, the recent observation
of charged $Z$-states, such as $Z_c(\num{3900})^\pm$ and
$Z_b(\num{10610})^\pm$, has proven the existence of mesons with exotic
structure~\cite{Liu:2013dau,Ablikim:2013wzq,Belle:2011aa}.  Already
the existence of one system with a wave function and/or quantum
numbers requiring an explanation beyond \enquote{ordinary} mesons
implies the existence of a large number of additional matter states.
Unless new principles for the construction of color-neutral hadrons
are found, this should hold for all quark-flavour combinations.

In the sector of light mesons, the issue of exotic states remains
unresolved.  The lowest-mass state discussed in this context is the
\PfZero, which contains \nnbar ($n = \{u, d\}$) and a dominant \ssbar
component.  It has also been interpreted as a \KKbar
molecule~\cite{Jaffe:1976ig,Hanhart:2007wa,Tornqvist:1995kr}.  The
\PfOne[1420] with a width of only \SI{55}{\MeVcc} couples strongly to
\KKstarbar and was also suggested to be a molecular-type
structure~\cite{Longacre:1990uc}.  In \refCite{Beringer:1900zz}, the
Particle Data Group has tentatively adopted the scenario of
\PfOne[1420] being the SU(3) partner of \PfOne[1285].  In the class of
spin-exotic mesons, the \PpiOne[1600] is the only meson observed by
several experiments in different decay modes.  However, the masses
quoted and in particular the widths vary considerably between
different experiments, and values for the width often exceed
\SI{400}{\MeVcc}.  This suggests the existence of dynamical effects
similar to the case of \PaOne.  The situation is characterized by
individual states without recognizable pattern and, except for \PaZero
and \PfZero, the absence of any observed isospin partners.

In order to search for new exotic mesons, we have analyzed the
diffractive reaction \reaction with the focus on waves with
quark-model \JPC combinations~\footnote{The $C$-parity refers to the
neutral state of the isospin multiplet.}.  We have studied the
$\JPC = \onePP$ states in order to search for a possible partner of
the isosinglet \PfOne[1420].  Our analysis aims at the charged isospin
$I = 1$ analogue decaying into \threePi.  Although this final state
and the mass range of \SIrange{1}{2}{\GeVcc} have already been studied
by many experiments, see
\eg~\refsCite{daum:1980ay,amelin:1995gu,Chung:2002pu}, the improvement
by almost two orders of magnitude in sample size has opened a new
avenue for analysis.
 %
%
%

The COMPASS experiment~\cite{Abbon:2007pq,Abbon:2014aex} is located at
the M2~beam line of the CERN Super Proton Synchrotron.  For this
measurement, we used a beam of \SI{190}{\GeVc} $\pi^-$ with
\SI{96.8}{\percent} purity, impinging on a \SI{40}{\cm} long
liquid-hydrogen target that was surrounded by a recoil-proton detector
(RPD).  Incoming pions were identified with a Cherenkov counter placed
in the beam line at the entrance to the experimental area.  The
large-acceptance high-precision spectrometer is well suited for
investigating high-energy reactions at low to intermediate values of
\tpr, which denotes the reduced squared four-momentum transfer to the
target proton with $\tpr \equiv \tabs - \tmin$.  For this measurement,
\tpr is chosen to be in the range from \SIrange{0.1}{1.0}{\GeVcsq},
where the lower bound is dictated by the acceptance of the RPD and the
upper bound by the steep drop of the number of events with increasing
\tpr.  Outgoing charged particles are detected by the tracking system
and their momenta are determined using two large-aperture magnets.

The data presented in this Letter were recorded in the year 2008.  A
detailed description of setup, data selection, and analysis scheme can
be found in \refsCite{COMPASS_3pi:2014,haas:2014bzm}.  The trigger is
based on a recoil-proton signal in the RPD in coincidence with an
incoming beam particle and no signal in the beam-veto counters.  We
require a production vertex located within the target volume, with one
incoming beam-pion track and three outgoing charged particles,
compatible with the pion hypothesis based on information from the RICH
counter.  The sum of the momenta of the outgoing particles is required
to be equal to the average beam momentum within two standard
deviations, \ie $\pm \SI{3.78}{\GeVc}$.  We require Feynman-$x$ of the
fastest final-state pion to be below~0.9 for rapidity differences
between the fast $\pi^-$ and the slower \twoPi pair in the range
\numrange{2.7}{4.5}.  This suppresses the small contamination of
centrally produced final states, which contribute mainly at higher
$3\pi$ masses.  A total of \num{46E6}~events was selected in the mass
range between \SIlist{0.5;2.5}{\GeVcc}.
 %
%
%

In order to extract the spectrum of resonances produced in the
reaction, we have performed a partial-wave analysis (PWA) that is
pursued in two steps.  First, we fit the intensity distributions in
the 5-dimensional phase space independently in one hundred
\SI{20}{\MeVcc} wide bins of $3\pi$ mass \mThreePi, each divided into
11~bins of \tpr.  We adopt the notation
\wave{J}{PC}{M}{\refl}{[\text{isobar}]}{L} to define partial waves.
Here, \refl denotes the reflectivity and $M \geq 0$ the magnitude of
the spin projection along the beam axis (see \refCite{Chung:1974fq}),
while $L$ is the orbital angular momentum between the isobar and the
bachelor pion in the decay of the $3\pi$ state.  We use the isobar
model, which for our fits contains 88~waves, \ie 80~waves with
positive and 7~with negative reflectivity as well as one
non-interfering wave representing three uncorrelated pions.  This set
contains all significant isobars that decay into \twoPi and has been
derived from a larger set with 128~waves by requiring a minimum
relative intensity of about \num{E-4}.  The likelihood fit function is
built from two incoherently added terms that correspond to the two
values of reflectivity, $\refl = \pm 1$.  Each term coherently sums
over all partial-wave amplitudes that belong to the respective value
of \refl.  Details on the fit model, the fitting procedure, and the
results are described in \refsCite{COMPASS_3pi:2014,haas:2014bzm}.
The division of our data set into 11~bins of \tpr is motivated by the
very different \tpr-dependences of resonant and non-resonant
components~\cite{daum:1980ay,COMPASS_3pi:2014}.  In all partial waves
studied, the intensity of non-resonant, \ie Deck-like
components~\cite{Deck:1964hm}, drops off much faster with increasing
\tpr than that of resonances.

\begin{figure*}
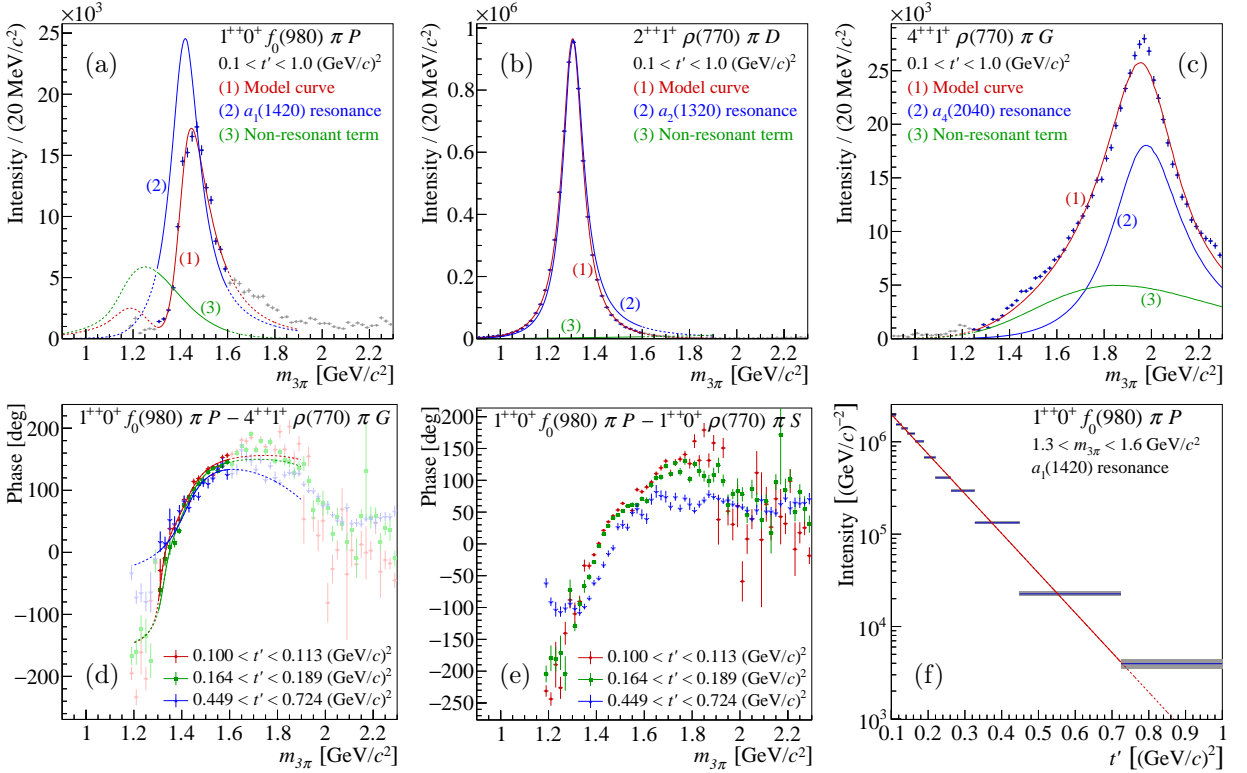

  \vspace*{-2ex}
  \centering
  \subfloat{\subfigimg[width=0.333\linewidth,pos=ul,hsep=2.75em]{(a)}{fig_1a}
    \label{fig:intensity_1pp_f}
  }
  \subfloat{\subfigimg[width=0.333\linewidth,pos=ul,hsep=2.75em]{(b)}{fig_1b}
    \label{fig:intensity_2pp_f}
  }
  \subfloat{\subfigimg[width=0.333\linewidth,pos=ur,hsep=1.00em]{(c)}{fig_1c}
    \label{fig:intensity_4pp_f}
  }
  \\[-3ex]
  \subfloat{\subfigimg[width=0.333\linewidth,pos=bl,hsep=2.75em,vsep=7ex]{(d)}{fig_1d}
    \label{fig:phase_a4}
  }
  \subfloat{\subfigimg[width=0.333\linewidth,pos=bl,hsep=2.75em,vsep=7ex]{(e)}{fig_1e}
    \label{fig:phase_a1_1260}
  }
  \subfloat{\subfigimg[width=0.333\linewidth,pos=bl,hsep=2.75em,vsep=7ex]{(f)}{fig_1f}
    \label{fig:t-dependence_a1_1420}
  }
  \caption{(Color online) Results of the PWA in $3\pi$ mass bins of
    \SI{20}{\MeVcc} width (data points with statistical errors only).
    The red curves in the intensity distributions [panels~(a)--(c)]
    represent the fit model, which is the coherent sum of resonances
    (blue) and non-resonant contributions (green).  The fit is
    constrained to the mass range indicated by the continuous curves.
    Extrapolations of the model are shown as dashed curves.
    Panel~(d)~shows the relative phase between $1^{++}$ and $4^{++}$
    together with the model curves and~(e) the phase between two
    $1^{++}$ decay modes. Here, the different colors distinguish
    3~exemplary \tpr~bins.  The \tpr-dependence of the \PaOne[1420]
    intensity is shown in~(f) together with the result of a
    single-exponential fit (red line) yielding a slope parameter of
    $b \approx \SI{10}{\perGeVcsq}$.  }
  \label{fig:massDepFit}
\end{figure*}

The result of this first PWA step reveals a number of well-known
resonances with $\JPC = 0^{-+}$, $2^{-+}$, $1^{++}$, $2^{++}$, and
$4^{++}$.  They are identified by structures in the mass spectra and a
mass-dependent phase that is measured against the reference wave
\wave{1}{++}{0}{+}{\Prho}{S}.  The \wave{1}{++}{0}{+}{\PfZero}{P}
intensity shows a clear signal slightly above \SI{1.4}{\GeVcc}, which
cannot be associated with a known $1^{++}$ state [see data points in
\cref{fig:intensity_1pp_f}].  Rapid phase rotations \wrt known
resonances are observed in the signal region, independent of \tpr [see
points in \cref{fig:phase_a4,fig:phase_a1_1260}].  The same feature is
observed in the \threePiN final state~\cite{Uhl:2014lva}.

In the second analysis step, we use a resonance model to fit the
spin-density matrices resulting from the first analysis step
simultaneously in all bins of \tpr and in wave-specific ranges in
\mThreePi.  Typically, only subsets of these spin-density matrices are
fit simultaneously.  In this Letter, we present such a fit using a
minimal set of 3~waves, namely \wave{2}{++}{1}{+}{\Prho}{D},
\wave{4}{++}{1}{+}{\Prho}{G}, and \wave{1}{++}{0}{+}{\PfZero}{P}.  The
first two waves contain the known \PaTwo and \PaFour.  These two waves
act as interferometers in order to search for structures in
\wave{1}{++}{0}{+}{\PfZero}{P}, where no resonances have yet been
reported.  For this fit, we model the amplitude of each wave by a
coherent superposition of a resonant contribution, which is described
by a relativistic Breit-Wigner~(BW) amplitude and a non-resonant
contribution.  Hence both components are allowed to interfere.  In the
$4^{++}$ and $1^{++}$ waves, the latter are described by terms of the
form $\mathcal{F}(\mThreePi) = e^{-c_1\, q^2(\mThreePi)}$, where $c_1$
is a fit parameter and $q$ is the two-body break-up momentum for a
particular isobar at the mass \mThreePi.  For the non-resonant term in
the $2^{++}$ wave, this parametrization is extended to include an
explicit \tpr-dependence.  A simple BW amplitude is used for the
\PaFour and the $\JPC = 1^{++}$ state, a BW amplitude with
mass-dependent width for the \PaTwo.  In the latter, the decay phase
space is approximated assuming quasi-two-body decays into
\SI{80}{\percent} $\Prho\,\pi$ and \SI{20}{\percent} $\Peta*\,\pi$.

The result of this fit is shown as curves in
\cref{fig:intensity_1pp_f,fig:intensity_2pp_f,fig:intensity_4pp_f,fig:phase_a4},
in which the model curves describe the data well. The blue curve in
\cref{fig:intensity_1pp_f} reveals the existence of a new axial-vector
state in the \wave{1}{++}{0}{+}{\PfZero}{P} wave, which we introduce
as \PaOne[1420].  This wave collects only \SI{0.25}{\percent} of the
total observed intensity.  Its resonance interpretation is supported
by the observation of a rapid phase variation by about
\SI{180}{\degree} across the peak region \wrt the $4^{++}$ [see
\cref{fig:phase_a4}] and $2^{++}$ reference waves.  As illustrated in
\cref{fig:phase_a1_1260}, the \wave{1}{++}{0}{+}{\PfZero}{P} wave
shows similarly rapid phase motion also relative to the
\wave{1}{++}{0}{+}{\Prho}{S} wave.  This indicates that the observed
structure in the $\PfZero\,\pi$ decay mode is not caused by the
high-mass tails of the \PaOne, which dominates the $\Prho\,\pi$ wave.
Our fit reveals a BW~mass for the \PaOne[1420] of \SI{1414}{\MeVcc}
and a width of \SI{153}{\MeVcc}.  The observed shift of the measured
peak position \wrt the resonance position in
\cref{fig:intensity_1pp_f} is due to destructive interference of the
BW resonance (blue) and the non-resonant term (green).  In this wave,
the fit model is chosen to cover the mass range up to
\SI{1.6}{\GeVcc}.  The observed tail of the intensity at higher masses
may be attributed to non-resonant contributions that are not described
by the present model.

The resonance-model fit is performed simultaneously in all 11~bins of
\tpr.  We allow production strengths and phases of resonances and
non-resonant contributions to vary with \tpr.  Spectral shapes and
BW~parameters are assumed to be independent of \tpr.
\Cref{fig:t-dependence_a1_1420} shows the resulting \tpr-spectrum of
the production intensity of the BW~amplitude that describes the
\PaOne[1420].  The BW intensity and that of the non-resonant
contribution show a steep, approximately exponential \tpr-dependence.
Fits with a single exponential yield the slope parameters.  Resonances
are typically described by slope parameters
$b \approx \SIrange{8}{10}{\perGeVcsq}$ that are smaller than those of
the non-resonant contributions with
$b \approx \SIrange{12}{15}{\perGeVcsq}$~\cite{COMPASS_3pi:2014}.  The
new \PaOne[1420] has a slope parameter of
$b \approx \SI{10}{\perGeVcsq}$ that is similar to those of \PaTwo and
\PaFour, which supports its resonance interpretation.  The fact that
the \PaOne[1420] is produced with nearly constant phase offset
relative to the \PaTwo, independent of \tpr, provides further support
for this interpretation.  As expected, the slope of the non-resonant
contribution in the $1^{++}$ wave is steeper with
$b \approx \SI{13}{\perGeVcsq}$.

The 88 partial-wave set contains three independent contributions for
the \pipiSW isobars, namely the \PfZero with parameters taken from
\refCite{Ablikim:2004wn}, a broad component denoted \pipiS taken from
elastic \pipiSW scattering~\cite{Au:1986vs}, and the \PfZero[1500]
described by a simple BW~amplitude.  The \PaOne[1420] is observed only
in \wave{1}{++}{0}{+}{\PfZero}{P}, while no signal with corresponding
phase motion is seen in \wave{1}{++}{0}{+}{\pipiS}{P} or any other
$1^{++}$ wave.  In order to confirm the unique coupling of
\PaOne[1420] to $\PfZero\,\pi$, we have investigated in a separate
study~\cite{COMPASS_3pi:2014,Krinner:2014zsa} the structure of the
\twoPi subsystem forming \zeroPP isobars using a novel fit procedure
for the partial-wave decomposition in bins of \mThreePi and \tpr.
Instead of describing the \zeroPP isobars by several amplitudes with
fixed shape, their mass dependence is replaced by a piecewise constant
function across \mTwoPi, which is determined from data.  We thus
remove possible bias originating from the isobar model for $0^{++}$.
For \mThreePi around the new resonance, a clear intensity correlation
of the new \PaOne[1420] with the \PfZero is observed within the
extracted \zeroPP isobar amplitude~\cite{COMPASS_3pi:2014}.

Due to the large data set, statistical uncertainties of the extracted
resonance parameters are negligible compared to systematic ones.
Therefore, we performed extensive systematic studies concerning
event-selection cuts, the model used in the first step of the PWA fit,
as well as wave set and parametrizations employed in the
resonance-model fit.  The result is stable under all these studies.
The main systematic uncertainties arise from the resonance-model fit.
The estimated total systematic uncertainty is
\Numaerr{15}{13}\valueSep\si{\MeVcc} for the \PaOne[1420] mass and
\Numaerr{8}{23}\valueSep\si{\MeVcc} for the width.  This was obtained
by changing the set of waves used in the resonance-model fit, the
event selection criteria, and the parametrization of the non-resonant
terms.  Using instead of a simple BW~amplitude one with a
mass-dependent width, with $\PfZero\,\pi$ as the only decay channel,
yields central values for mass and width of \SI{1433}{\MeVcc} and
\SI{146}{\MeVcc}, respectively.  Pseudo data simulated using the full
88 waves with and without the \wave{1}{++}{0}{+}{\PfZero}{P} wave,
which were subsequently analyzed with our standard 88-wave PWA, show
no indication for model leakage artificially populating the
\wave{1}{++}{0}{+}{\PfZero}{P} wave.

We have estimated the statistical significance of the \PaOne[1420]
signal by rescaling the statistical uncertainties of the data points
such that the fit has a $\chi^2 / \text{ndf} = 1$ (following
\refCite{Beringer:1900zz}).  The decrease of the $\chi^2$ probability
of a fit performed without the assumption of a new resonance in the
$1^{++}$ wave translates into a significance of the \PaOne[1420]
measurement, which by far exceeds the value of $5\,\sigma$.  The same
result is obtained when only those spin-density matrix elements are
included in the $\chi^2$ calculation, which contain the $1^{++}$ wave.
Therefore, the result does not depend on possible imperfections in the
description of the $2^{++}$ and $4^{++}$ reference waves at masses
much below or above the new \PaOne[1420].

Summarizing our analysis, we have performed a resonance-model fit
based on a spin-density submatrix that was obtained by the so far most
extensive 88-wave $3\pi$ PWA using the large COMPASS data set.
Restricting this resonance-model fit to the three waves
\wave{2}{++}{1}{+}{\Prho}{D}, \wave{4}{++}{1}{+}{\Prho}{G}, and
\wave{1}{++}{0}{+}{\PfZero}{P}, we have observed a new \PaOne* meson
at $m = \SIaerr{1414}{15}{13}{\MeVcc}$ with a width of
$\Gamma = \SIaerr{153}{8}{23}{\MeVcc}$.  This finding was made
possible by our large event sample and, when compared to previous
experiments, also by the more homogeneous acceptance of the COMPASS
setup in the five phase-space variables used in the
PWA~\cite{Abbon:2014aex,haas:2014bzm}.

The interpretation of this new state is still unclear.  Quark-model
calculations including tetra-quark states predict
$nn\,\overline{n}\overline{n}$ and $ns\,\overline{n}\overline{s}$
iso-vector states at \SIlist{1381;1530}{\MeVcc},
respectively~\cite{Vijande:2005jd}.  The molecular model for the
\PfOne[1420] proposed in \refCite{Longacre:1990uc} could possibly be
extended to the isovector case.  After our first announcement of the
\PaOne[1420]~\cite{StephanPaul:2013xra}, several explanations were
proposed~\cite{wang:2014bua,Basdevant:2015ysa,Basdevant:2015wma,Ketzer:2015tqa}, none of
them describing all features of the data.  The properties of the
\PaOne[1420] suggest it to be the isospin partner of the \PfOne[1420].
This is supported by its mass value of \SI{1414}{\MeVcc} and by its
strong coupling to \PfZero, which is interpretable as a \KKbar
molecule.  The \PaOne[1420] width of only \SI{153}{\MeVcc} is narrow
as compared to most other known iso-vector states, which typically
have widths between \SIlist{250;350}{\MeVcc}.  The much smaller width
of the \PfOne[1420] of only \SI{54.9 (26)}{\MeVcc} can be explained by
its strong coupling to \KKstarbar with the corresponding phase space
being much smaller than that for \PaOne[1420] decaying into
$\PfZero\,\pi$.  The \PaOne[1420] and the \PfOne[1420] may possibly be
the first observed isospin partners for a $\pi\KKbar$ molecular-type
excitation, which obey isospin symmetry.  This interpretation suggests
further experimental and theoretical studies of the $\pi\KKbar$ final
state.
 %
%
%

We gratefully acknowledge the support of the CERN management and staff
as well as the skills and efforts of the technicians of the
collaborating institutions.  This work is supported by MEYS (Czech
Republic); ``HadronPhysics2'' Integrating Activity in FP7 (European
Union); CEA, P2I and ANR (France); BMBF, DFG cluster of excellence
``Origin and Structure of the Universe'', the computing facilities of
the Computational Center for Particle and Astrophysics (C2PAP),
IAS-TUM and Humboldt foundation (Germany); SAIL (CSR) (India); ISF
(Israel); INFN (Italy); MEXT, JSPS, Daiko and Yamada Foundations
(Japan); NRF (Rep. of Korea); NCN (Poland); FCT (Portugal); CERN-RFBR
and Presidential Grant NSh-999.2014.2 (Russia).

%
\providecommand{\href}[2]{#2}\begingroup\raggedright\endgroup
 %


\end{document}